\documentclass[conference]{IEEEtran}
\IEEEoverridecommandlockouts

\usepackage{amssymb}
\usepackage{graphicx}
\usepackage{qtree}
\usepackage{circuitikz}
\usepackage{tikz}
\usepackage{mathtools}
\usepackage{booktabs}
\usepackage{hyperref}
\usepackage{subcaption}
\usepackage{dirtree}
\usepackage{booktabs}
\usepackage{multirow}

\usetikzlibrary{shapes,snakes}

\usepackage{amsthm}

\theoremstyle{definition}

\usepackage[linesnumbered,ruled,vlined]{algorithm2e}
\usepackage{algpseudocode}

\usetikzlibrary{shapes.gates.logic.US,trees,positioning,arrows}

\usepackage{lipsum}

\usetikzlibrary{shapes,arrows}

\usepackage{pgfplots}
\pgfplotsset{width=8.3cm,compat=1.9}

\usepackage{pgfplotstable}
\usepackage{pgfplots}
\usetikzlibrary{patterns}
\pgfplotsset{compat=1.3}
\usepackage{xcolor}
\usepackage{csvsimple}

\usepackage[linesnumbered,ruled,vlined]{algorithm2e}
\usepackage{algpseudocode}

\usepackage{amssymb}
\usepackage{graphicx}
\usepackage{qtree}
\usepackage{circuitikz}
\usepackage{tikz}
\usepackage{mathtools}
\usepackage{booktabs}
\usepackage{hyperref}
\usepackage{subcaption}
\usepackage{dirtree}
\usepackage{multirow}

\usetikzlibrary{shapes,snakes}

\usepackage{amsthm}

\usepackage[linesnumbered,ruled,vlined]{algorithm2e}
\usepackage{algpseudocode}

\usetikzlibrary{shapes.gates.logic.US,trees,positioning,arrows}

\usepackage{mathtools,amssymb,lipsum}

\usepackage{cuted}
\usetikzlibrary{shapes,arrows}

\usepackage{pgfplots}
\pgfplotsset{width=8.3cm,compat=1.9}

\usepackage{pgfplotstable}
\usepackage{pgfplots}
\usetikzlibrary{patterns}
\pgfplotsset{compat=1.3}
\usepackage{xcolor}
\usepackage{csvsimple}

\begin{document}

\title{A Markov Game Model for AI-based Cyber Security Attack Mitigation}

\author{\IEEEauthorblockN{Hooman Alavizadeh\textsuperscript{1}, Julian Jang-Jaccard\textsuperscript{1}, Tansu Alpcan\textsuperscript{2} and Seyit A. Camtepe\textsuperscript{3}}
		\IEEEauthorblockA{\textsuperscript{1} School of Natural and Computational Sciences,\\ Massey University, Auckland, New Zealand\\
		\textsuperscript{2} Department of Electrical and Electronic Engineering,\\ University of Melbourne, Parkville, Australia\\
		\textsuperscript{3} CSIRO Data61, Australia\\
		Email: \{h.alavizadeh, j.jang-jaccard\}@massey.ac.nz}
}

 \maketitle
	\begin{abstract}
    The new generation of cyber threats leverages advanced AI-aided methods, which make them capable to launch multi-stage, dynamic, and effective attacks. Current cyber-defense systems encounter various challenges to defend against such new and emerging threats.
Modeling AI-aided threats through game theory models can help the defender to select optimal strategies against the attacks and make wise decisions to mitigate the attack's impact.
	This paper first explores the current state-of-the-art in the new generation of threats in which AI techniques such as deep neural network is used for the attacker and discusses further challenges.
We propose a Markovian dynamic game that can evaluate the efficiency of defensive methods against the AI-aided attacker under a cloud-based system in which the attacker utilizes an AI technique to launch an advanced attack by finding the shortest attack path. We use the CVSS metrics to quantify the values of this zero-sum game model for decision-making.
	\end{abstract}


\section{Introduction}
A strong cyber defense system should be able to defend against the new generation of cyber threats~\cite{hou2020low,brundage2018malicious}. AI-powered threats are emerging attacks that use AI capabilities to launch various types of attacks to the system such as targeted attacks, adaptive attacks~\cite{hu2015dynamic}, DeepLocker~\cite{stoecklin2018deeplocker}, and so forth~\cite{wang2020ai}.


Although the traditional adversarial attacks were relatively easier to detect and defend as the attack patterns were algorithmic, the emerging attacks leverage AI features such as machine learning (ML) and deep learning to make malware much more evasive and pervasive. 
In addition, the AI techniques combined with automation can make the malware more scalable and make them able to attack the targets without human intervention~\cite{wang2020ai}. 

Game theory is a mathematical model based on a decision-responsive manner that makes a variation on strategies for each player according to the decision or movements of other players. In fact, game theory can be defined as studying the cooperation and conflict between rational and intelligent decision-makers based on the principles of using mathematical models. Game Theory models have been extensively studied for their use in cyber security and have shown to be very effective in the evaluation of defensive systems and addressing the security of networks and systems~\cite{manshaei2013game,attiah2018game,rass2017physical}.


In this paper, we consider an advanced attacker which is able to utilize AI techniques such as deep neural network (DNN) to find their target in a networked system fast and effectively. We assume that the attacker can leverage deep learning techniques to estimate the shortest attack path in a networked model (such as those proposed in \cite{rizi2018shortest,salehi2020shortest,qu2013efficient}). This enables the attackers to be resilient against dynamic defense techniques~\cite{cho2020toward,alavizadeh2019automated}. Moreover, we show how game theory can be leveraged to evaluate the dynamic defensive scenarios for avoiding these kinds of AI-powered attacks.
 
Concretely, we address the specific problem in which the AI-aided attacker can find the shortest attack path in a cloud. The cloud system and attack model are modeled as a zero-sum Markov Game. The cloud model is based on an Attack Graph (AG) representing the states of the game.
We define a zero-sum Markovian game model that captures the capabilities of the attacker and defender. In this case, the attacker can choose the actions that exploit the real-world vulnerabilities reported in the Common Vulnerabilities and Exploits (CVEs) through National Vulnerability Database (NVD). Then, the defender's actions are modeled based on a dynamic defense consisting of the placement of detection systems (such as IDS) to the hosts in the cloud. We design the rewards of this game by leveraging the Common Vulnerability Scoring Systems (CVSS) values such as attack impact and exploitability. This helps the defender to select appropriate strategies using the limited number of monitoring actions for each state of the game. The defender action is the placement of an IDS in a host in the cloud to monitor and detect any prospective threats. However, the defense action may incur some performance degradation and cost to the defender.
 
In this paper, we utilized a Markovian game model to analyze AI-aided attacks for a cloud model which can help the defender to mitigate attacker rewards using a dynamic IDS placement strategy. The main contributions of this paper are as follows:

\begin{itemize}
    \item We provide an up-to-date review on the new generation of threats such as AI-powered threats and categorize them based on two main threat types which are \textit{AI-aided attacks} and \textit{AI-embedded attacks}.
	\item We discuss the AI-aided attack approaches such as those using graph-based AI techniques to find the targets in a networked system effectively. We investigate in depth a threat model which is able to utilize Deep Neural Network (DNN) to find the shortest attack path in a cloud to launch the attack efficiently.
    \item We propose a Markovian game model that can evaluate the effectiveness of the state-of-art defense mechanisms against the AI-aided attacks in which the attacker utilizes an AI technique such as DNN to estimate the shortest attack path in a cloud system. We quantify the game parameters based on a zero-sum game and CVSS values.
	\item We offer the formal mathematical definitions for the proposed game model. We also determine the probabilistic values of the states of the game. Finally, we clearly formalize, analyze, and quantify actions and states transition probabilities for the game model.
\end{itemize}

The rest of the paper is organized as follows. Section \ref{sec:back} discusses some essential concepts which are related to this paper including AI-powered threats categorization. The related work is given in Section~\ref{sec:rel}. In Section~\ref{sec:proposed}, we define the necessary concepts, definitions, mathematical notations, and propose our Markovian game model. Discussion and limitations of the current study are presented in Section \ref{sec:discussion}. Finally, we conclude the paper in Section~\ref{sec:conclusion}.

\section{AI-powered Threats}\label{sec:back}

\begin{figure}[t]
	\centering
	\includegraphics[width=0.95\linewidth]{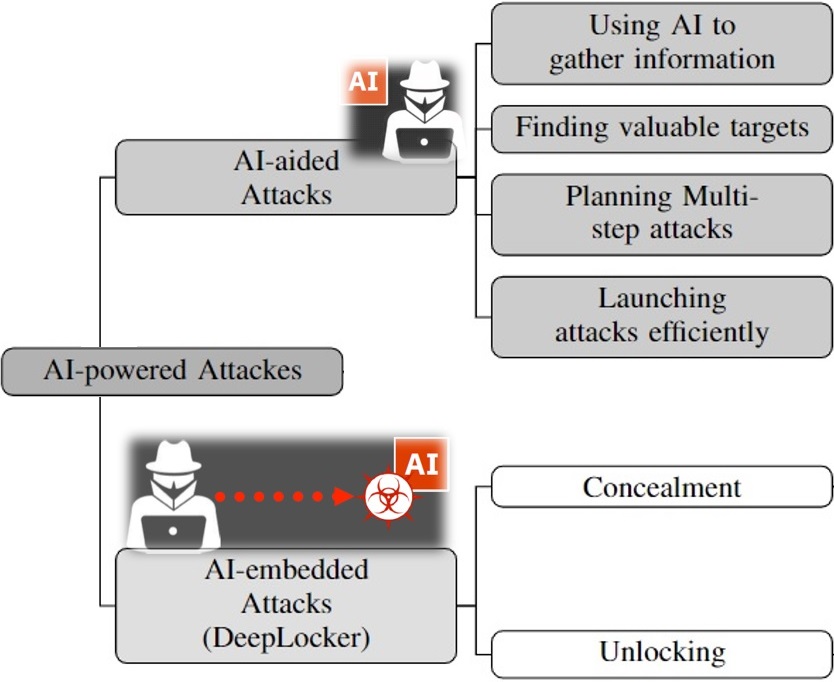}
	\caption{AI-powered threats categorization.}
	\label{fig:AI-pow}
\end{figure}

{In these days, attackers use innovative and smart methods for launching various types of attacks. It includes delivering malicious activities, exploiting zero-day vulnerabilities, and use of deep neural networks to find the targets effectively. Thus, the cyber defensive system should be able to deal with a wide range of more intelligent, persistent, and sophisticated attacks equipped with more advanced technologies such as AI-powered attacks. In~\cite{kaloudi2020ai}, the authors investigated the AI-powered cyber attacks and mapped them onto a proposed framework with new threats including the classification of several aspects of threats that use AI during the cyber-attack life cycle. We categorize the AI-powered attacks based on {\em AI-aided} and {\em AI-embedded} attacks as shown in Figure~\ref{fig:AI-pow}. AI-aided attacks are of those who leverage AI to launch the attacks effectively. In this type, the intelligent attackers use AI techniques. However, in AI-embedded attack, the threats are weaponized by AI themselves such as Deep locker~\cite{stoecklin2018deeplocker}.}

In AI-aided attacks, the attackers use AI to launch various attacks based on different targets such as adaptive attacks or multi-stage attacks which need knowledgeable attackers equipped with prior knowledge of the target system. Moreover, adaptive attacks are known as intelligent attacks. In these types of attacks, the {attacks can be adapted to the dynamic environment that changes the system's conditions. These types of attackers are intelligent enough so that they can wisely manage their resource limitations while they opportunistically try to compromise the entire system such as executing adaptive attacks~\cite{tramer2020adaptive}.}
The attacker could launch various AI-based techniques to detect and recognize the target network, vulnerabilities, and valuable targets~\cite{kaloudi2020ai}. Moreover, the new type of AI-aided attackers are able to use high intelligence and sufficient resources to  launch increasingly more intelligent, sophisticated and persistent attacks. This enables the attackers to find valuable targets in the network, find attack path efficiently, and obtain highly sensitive information~\cite{tramer2020adaptive,hutchins2011intelligence}. In particular, attackers are intelligent enough to (i) optimize the attack; for instance, they can estimate the most efficient attack path in a networked system with lower cost and higher benefits~\cite{feng2017signaling}, (ii) execute multi-stage attacks including reconnaissance phase by scanning the system prior to real attacks exploiting the system's components (such as virtual machines). They are able to continue to the attack by exploiting other system's components after they break into the system~\cite{hu2015dynamic}.

\begin{table*}[t]
\centering
\footnotesize
\caption{Comparison of Graph-based techniques which can be leveraged by attackers to find attack path such as shortest attack path in the network}
\begin{tabular}{@{}|l|l|l|l|l|l|l|@{}}
\hline
\textbf{Approach}                                                                              & \textbf{Technique}                                               & \textbf{Advantages}                                                                       & \textbf{Disadvantages}                                                                                    & \textbf{Practicality} & \textbf{\begin{tabular}[c]{@{}l@{}}Resilience to\\ dynamic defense\end{tabular}} & \textbf{Ref.}       \\ \hline
\multirow{2}{*}{\textbf{\begin{tabular}[c]{@{}l@{}}Classic search \\ strategies\end{tabular}}} & BFS/DFS                                                          & Fast for small size                                                                 & \begin{tabular}[c]{@{}l@{}}--Not scalable\\ --Time complexity\\ --Exhaustive search\end{tabular} & High                  & Verylow                                                                          & \cite{bagheri2008finding} \\ \cline{2-7} 
                                                                                               & Dijkstra                                                         & \begin{tabular}[c]{@{}l@{}}--Fast for small size\\ --Consider priority\end{tabular} & \begin{tabular}[c]{@{}l@{}}--Not scalable\\ -- Resource consumption\end{tabular}                 & High                  & Low                                                                              & \cite{deng2012fuzzy} \\ \hline
\textbf{Traditional AI-based}                                                                  & $A^*$                                                            & Good performance                                                                    & \begin{tabular}[c]{@{}l@{}}--Needs heuristic\\ --Real implementation\end{tabular}                & Low                   & Low                                                                              & \cite{goldberg2005computing} \\ \hline
\multirow{3}{*}{\textbf{Learning-based AI}}                                                    & Machine Learning                                                 & Scalable                                                                            & Need Training                                                                                    & Medium                & High                                                                             & \cite{kojic2006neural} \\ \cline{2-7} 
                                                                                               & Neural Network                                                   & \begin{tabular}[c]{@{}l@{}}--Scalable\\ --Adaptive\end{tabular}                       & Need Training                                                                                    & Medium                & High                                                                             & \cite{kojic2006neural} \\ \cline{2-7} 
                                                                                               & Deep Learning                                                    & \begin{tabular}[c]{@{}l@{}}--Scalable\\ --Adaptive\end{tabular}                     & Need Training                                                                                    & Medium                & Very High                                                                        & \cite{rizi2018shortest,salehi2020shortest} \\ \hline
\textbf{\begin{tabular}[c]{@{}l@{}}Evolutionary\\ Computation (EC)\end{tabular}}               & \begin{tabular}[c]{@{}l@{}}Genetic Algorithm\\ (GA)\end{tabular} & Fast approach                                                                       & \begin{tabular}[c]{@{}l@{}}--Low scalability\\ -- not optimal\end{tabular}                       & Low                   & Medium                                                                           & \cite{bagheri2008finding,syarif2018solving} \\ \hline
\end{tabular}
\end{table*}

In contrast, the AI capabilities are embedded into the malware or threat itself in the AI-embedded attacks. Such as targeted attacks and Deep Locker~\cite{stoecklin2018deeplocker} in which the current defensive technologies cannot easily detect and defend against those types of attacks. Deeplocker uses neural network technologies and includes three layers of concealment which make the Deeplocker evasive to detect. In the first layer, the target is concealed (who and what the target is). The second layer is the target instance concealment which hides the specifics. Finally, the third layer conceals the malicious intent which encrypts the payload and hides the attack technique~\cite{stoecklin2018deeplocker}. Another example of AI-embedded attacks is the new generation of botnet attacks which are very difficult to be detected by the current decisive strategies~\cite{wang2020ai}. They utilized deep neural networks such as convolutional neural networks (CNN) to establish a covert channel between bots and botmasters. 

However, in this paper, we only consider AI-aided attacks in which the attacker can use a deep neural network to find out the shortest attack path in a targeted network.

\section{Related Work} \label{sec:rel} 
\subsection{Graph-based AI-aided Threats}
The application of Artificial Intelligence (AI) such as Genetic Algorithms and machine learning in the graph theory-based problems such as finding the fastest, shortest, or cheapest path in a network has been studied in various research, the application of Genetic Algorithm~\cite{bagheri2008finding}, Neural Network~\cite{kojic2006neural}, Deep Learning~\cite{rizi2018shortest,salehi2020shortest} and so on. However, the application of graph theory-based and learning-based approaches for finding the shortest and most suitable paths in a network can be leveraged by attackers to launch their attacks effectively. Finding the attack path from source to target in a graphical attack model is an important capability for the attackers~\cite{zimba2019bayesian,liu2020network}. This enables the attacker to reach the target with minimum effort and cost. 


\subsubsection{Classical Search Strategies}
{The classic algorithms for calculating shortest paths are Breadth-first search (BFS), Dijkstra's, and Bellman-Ford. Dijkstra's is the most used case of finding the shortest path between two specific nodes with no available path cost heuristic.} {The earliest work on neural-based solutions to the shortest path was motivated by communications and packet routing, where approximate methods faster than the classical algorithms were desired. These operate quite differently from today's neural networks, they used iterative back-propagation to solve the shortest path on a specific graph~\cite{kojic2006neural}}. {Groundbreaking is another approach to the problem was DeepMind's Differentiable neural computers, which computed shortest paths across the large graph of geographical places. The method is by taking the graph as a sequence of connection tuples and learning a step-by-step algorithm using a read-write memory. It was provided with a learning curriculum, gradually increasing the graph size.} However, the traditional methods have scalability problem for larger graphs. For instance, an efficient implementations of Dijkstra can compute the shortest paths for a node to others in $O(n log n+m)$ time in a graph with $n$ nodes and $m$ edges.

\subsubsection{Traditional AI-based Strategies}
The $A^*$ search algorithm uses heuristic-based techniques to compute the shortest path (it also is known as light generalization of Dijkstra algorithm). Practically, the $A^*$ algorithm as quickly as Dijkstra's algorithm. However, those methods suffer from high time complexity and difficulties in finding the appropriate heuristic. There are many methods to obtain the shortest attack path in a network such as the classical Dijkstra algorithm and traditional AI-based such as $A^*$. However, from the attacker's point of view, the shortest attack path has to be computed within a very short period of time and scalable manner in order to support time-constrained attacks which can be detected and deceived by asynchronous defensive mechanisms which change the attack surface periodically~\cite{cho2020toward}. However, attackers prefer to use more intelligent approaches to find their targets using the shortest attack path~\cite{zimba2019bayesian}. Many studies investigated the optimized method for finding the shortest path from the attacker's perspective. For instance, in \cite{zimba2019bayesian}, authors proposed an optimal algorithm for APT attackers which enables the attacker to find the shortest attack path from multiple sources in a graphical attack model using a Bayesian network. Moreover, in~\cite{bagheri2008finding}, the authors proposed an approach to find the shortest path using a learning-based Genetic Algorithm (GA) based on principles of stochastic search and optimization techniques which is a fast algorithm with minimum failure ratio. However, most of the classical search and traditional methods are not able to satisfy the essential requirements for high-speed and fast response in large-scale applications, adaptation with dynamic graphs, real-world problems, and so on. Moreover, finding global optimal solutions is difficult in these approaches while it is easier to find global solutions in the classical shortest path approaches and pulse-coupled neural network (PCNN)~\cite{huang2017time}.

\subsubsection{Learning-based AI approaches}
To address the drawbacks of traditional methods, some neural network architectures have been designed for finding shortest path. In fact, neural network can be used to either find or estimate the shortest path on a certain network effectively~\cite{huang2017time,qu2013efficient,salehi2020shortest}. For instance, in~\cite{qu2013efficient}, a dynamic algorithm is developed to compute shortest path for large-scale networks through a modified model of pulse-coupled neural networks (M-PCNNs). However, their method could be ineffective when multiple link changes occur. Thus, this method is not useful for the attackers when the defender makes the system dynamic by changing the attack surface. {In \cite{huang2017time}, the authors proposed the time-dependent shortest path problem which is able to find the globally optimal solution through a time-delay neural network (TDNN) framework}. Moreover, in~\cite{salehi2020shortest}, {the authors proposed a novel method that is able to estimate the nodes' distances through embedding graphs into an embedding space. In the proposed method, they used leveraged a feed-forward neural network which used vectors obtained from two recent graph embedding techniques. Finally, The method can produce the shortest distances for the majority of node pairs.}

\subsection{Game Theory Review}
This section provides an overview of research on security defensive methods that use game-theoretic approaches. We present a selected set of works to highlight the application of game theory in addressing different forms of security-related problems. {Game theory has long been applied to study network security~\cite{alpcan2019decision,manshaei2013game}. Regarding the use of Bayesian game models, Alpcan and Basar~\cite{alpcan2006intrusion} proposed a security game between attacker and intrusion detection system (IDS) in the sensor network. They modeled their solution based on a finite Markov chain, Markov decision process, and Q-learning. 
Sheyner et al~\cite{jha2002two} presented a formal cost-benefit analysis for the attacks on a given network equipped with security measures for defending on the network attacks. In~\cite{chowdhary2016sdn}, the authors provided a method for attack graph construction based on network reconfiguration through parallel computing method. The proposed method can leverage the strategic reason information of attacks in large-scale systems.
}

Alpcan \& Basar~\cite{alpcan2003game}, proposed a two-player non-cooperative and non-zero-sum game to address the attack defense problem in the sensor networks. In their proposed game model, they assumed that the players have complete information about the system and the payoff functions of each other based on each player's optimal strategy. However, the drawback of the proposed game model is that the players have complete information about the game. Consequently, various relevant research introduced the game theory and studied the optimal strategy of network attack and defense. Moreover, in~\cite{jiang2009evaluating}, the authors proposed the offensive and defensive game model which utilized a network optimal active defense method. They solved a Nash equilibrium condition between the the attacker and defender to find the optimal attack and defense strategy. However, their main idea is to address the network security based on reducing the loss of security loss and risk management. In~\cite{gang2014network}, the authors proposed a non-cooperative non-zero-sum game model based on network security decision-making method for obtaining the optimal attack and defense evaluation. The proposed technique is able to generate an optimal strategy for attacker and defender by analyzing the interactions of both attackers and defenders. In~\cite{zhang2018network}, the authors used the offensive and defensive differential game and proposed a network security defense decision-making method. Based on their security evolution model, it analyzes the security state changing process of the network system and consequently generates the attack and defense differential game model. The proposed technique could provide the optimal defense strategy selection through the saddle point strategy selection. However, in the proposed game models, there is a lack of evaluation of more intelligent attackers based on the game models for defender and intelligent attackers.

\section{Threat Model} \label{sec:sp-DNN}
In this section, we demonstrated how a deep neural network can be used to estimate the shortest path in a graph efficiently. However, this model can be used by adversaries to learn the targeted system and launch a fast and effective attack toward the system. In \cite{salehi2020shortest}, the authors proposed a novel method for estimation of the shortest path distance between two nodes through vector embedding generated by deep learning approaches. The procedure undergoes three main phases:

\begin{itemize} 
    \item \textit{Graph analysis and embeddings}: The first step is to learn the nodes' mapping to a low dimension space of features maximizing the likelihood of preserving network neighborhoods of nodes using a graph embedding technique~\cite{grover2016node2vec}. An approach that is used for graph embedding is called the encoder-decoder approach which can encode the graph based on each node's attributes to a low-dimension vector and then decode the graph and related information from the low-dimension learned embeddings. However, graph embeddings include all required information for downstream machine learning tasks. The encoding function can be formulated as $Enc: V\rightarrow R^d$ based on~\cite{hamilton2017representation}, where $z_i$ denoted as the embedding for the node $v_i \in V$. Then, the decoder is defined as a function accepting a set of node embeddings and consequently decoding a user-specified graph statistics from the embeddings such that $Dec: R^d \times R^d \rightarrow R^+$. 
    \item \textit{Training set collection}: In this step, the training set can be achieved by computing the actual shortest path distances from a group of nodes to all of the remaining nodes. 
{The graph can be formally defined as $G(V, E)$ with $n$ nodes and $m$ edges. Based on two nodes $s$, $t \in V$, it can be defined as $ns$,$t=\{s, u1, u2, ...,ul-1, t\}$ to be a path of length $|ns,t|=l$ between $s$ and $t$, if $\{u1, ..., ul\}$ $\in V$ and ${(s, u1),(u1, u2), . . . ,(ul-1, t)} \in E$, and let $n_{s,t}$ be the set of all paths from $s$ to $t$. Moreover, let define $dG(s, t)$ as the shortest path length between any two nodes $s,t \in V$.} Then, a graph embedding technique described before can produce a vector embedding for the graph $G$ for every node $v \in V$ as $\theta(v)\in R^d$. To collect the training samples, the training pairs of the entire graph $G$ need to be extracted which computes the actual shortest distances from each landmark (group of nodes which denotes as $l$) to all of the remaining nodes using BFS search. It finally yields $l(n-l)$ training pairs. However, for testing pairs, the same strategy as training pairs can be used through considering a smaller set of landmarks and performing BFS traversals from landmarks to the other nodes which can generate a set of unseen pairs.
\item \textit{Neural Network Training}: The vectors of the training set (network embedding vector) will be fed into the feed-forward neural network (FNN) to estimate the distance between the nodes. Given a training pair $<\theta(v),\theta(U)>$, a joint representation as the input to the neural network can be generated through applying the binary operations such as concatenation, subtraction, point-wise multiplication, and average over the vector embeddings. The FNN consists of an input layer, a hidden layer, and an output layer and the size of the input layer depends on the binary operation on vector embeddings. Finally, the real-valued distance can be obtained by the neural network mappings of the input vectors to an output.
\end{itemize}

\begin{figure}[t]
	\centering
	\includegraphics[width=1\linewidth]{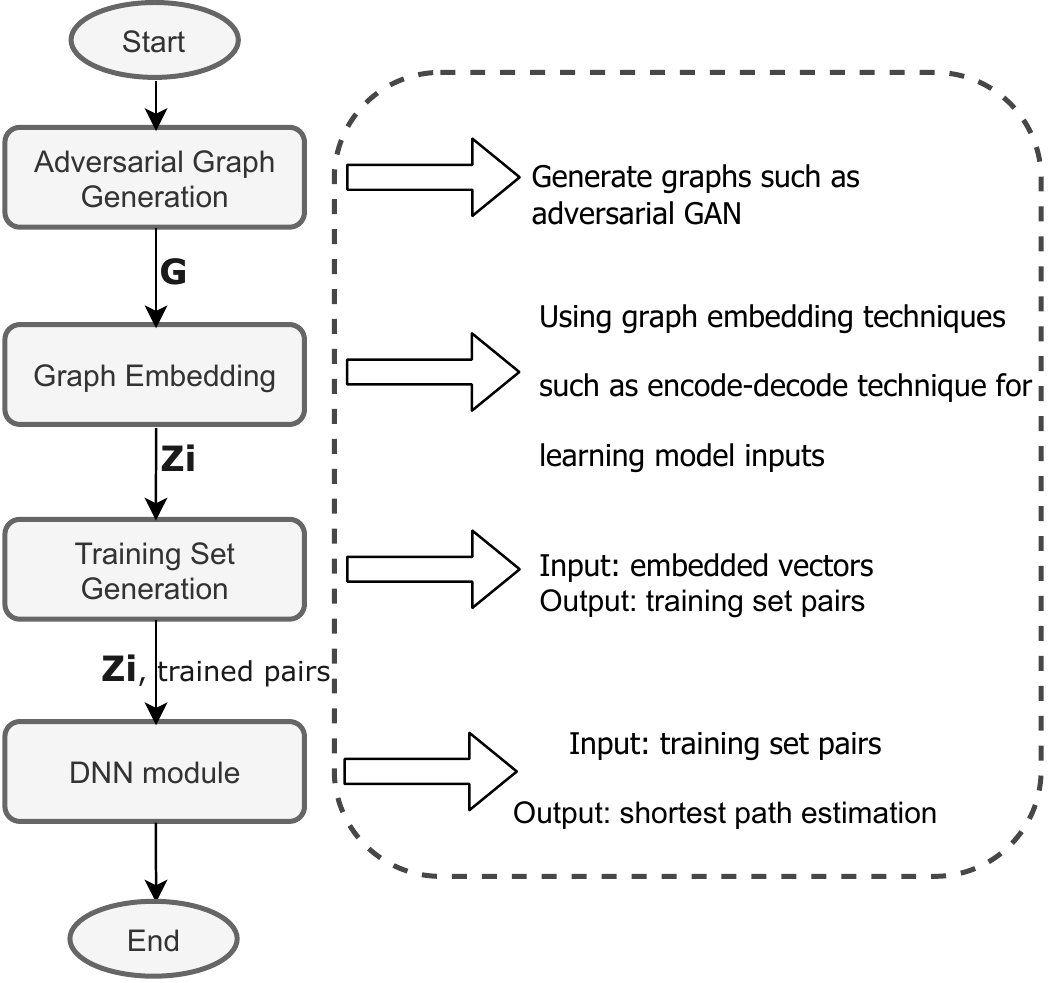}
	\caption{Shortest path estimation using Deep Neural Network steps.}
	\label{fig:steps}
\end{figure}

However, as demonstrated in Figure~\ref{fig:steps}, the advanced attackers can utilized AI techniques such as those as discussed above to find their target effectively and fast and can efficiently leverage deep learning techniques for finding the shortest path in a graph (such as those proposed in \cite{rizi2018shortest,salehi2020shortest,qu2013efficient}) which is resilience against the dynamic defense and also can provide the scalability of the model such as large cloud or enterprises. In this paper, we demonstrate how game theory can be leveraged to evaluate the defensive scenarios for avoiding these kinds of AI-powered attacks. 


\section{Game Theory for AI-aided Attacks and Dynamic Defence Evaluation}\label{sec:proposed}
In this section, we propose the system model for both attack and defense including the capabilities of AI-aided attackers which make them able to effectively find the shortest attack path in the modeled system. We then define and propose a zero-sum dynamic game model which can capture and evaluate various attacker actions and defenders in different discrete states of the game.

\subsection{Preliminaries}\label{sec:pre}
In this section, we introduce the notation and definitions used throughout this paper including the cloud model, attack, model, game theory definition, zero-sum Markovian game model, and so forth.


\begin{figure*}[t]
	\centering
	\begin{subfigure}{0.49\textwidth}
		\centering
    	\includegraphics[width=1\linewidth]{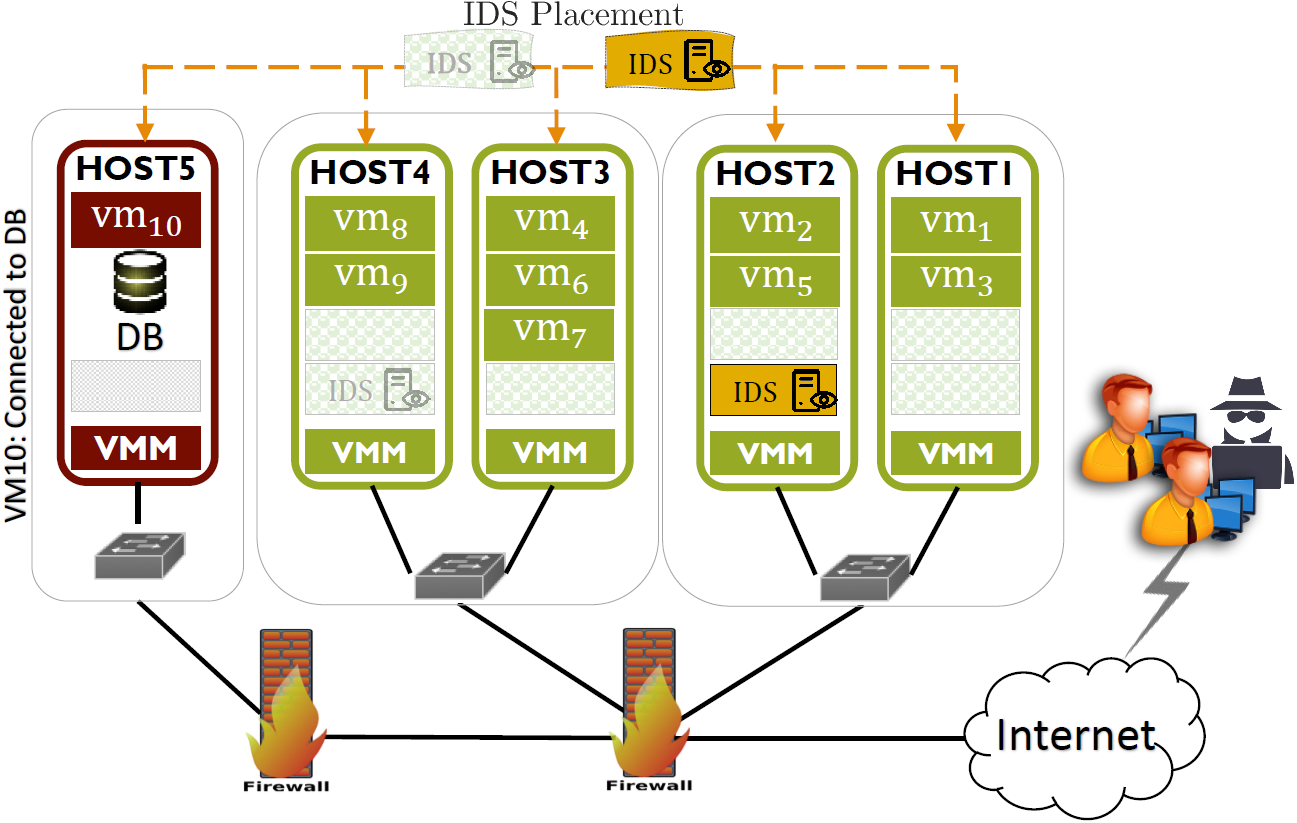}
		\caption{} 
		\label{fig:net}
	\end{subfigure}
	\begin{subfigure}{0.49\textwidth}
	\begin{tikzpicture}[scale=0.82, every node/.style={transform shape}]
	\node[shape=circle,draw=black,align=center,fill=red!10]  (A) at (8.6,0) {$A$};
	\node[shape=circle,draw=black,align=center] (v1) at (7,2) {$\text{h}_1$};
	\node[shape=circle,draw=black,align=center] (v2) at (7.7,-2) {$\text{h}_2$};
	\node[shape=circle,draw=black,align=center] (v4) at (6,0) {$\text{h}_4$};
	\node[shape=circle,draw=black,align=center,fill=yellow!10] (v5) at (5.5,-2.3) {$\text{h}_5$};
	\node[shape=circle,draw=black,align=center] (v3) at (4.4,2.4) {$\text{h}_3$};
	\node[shape=circle,draw=black,align=center] (v7) at (4.2,-1.1) {$\text{h}_7$};
	\node[shape=circle,draw=black,align=center,fill=yellow!10] (v6) at (2.8,1.1) {$\text{h}_6$};
	\node[shape=circle,draw=black,align=center,fill=blue!10] (v9) at (2.1,-1.8) {$\text{h}_9$};
	\node[shape=circle,draw=black,align=center] (v8) at (0.85,1.8) {$\text{h}_8$};
	\node[shape=circle,draw=black,align=center,fill=red!10] (v10) at (-0.5,-0.2) {$DB$};

	\path [->] (A) edge node[sloped,above,font=\scriptsize] {$e=0.53$} (v1);
	\path [dashed,line width=0.9pt,red,->] (A) edge node[sloped,above,font=\scriptsize] {$e=0.55$} (v2);
	\path [->] (v1) edge node[sloped,above,font=\scriptsize] {$e=0.51$} (v3);
	\path [->] (v1) edge node[sloped,above,font=\scriptsize] {$e=0.49$} (v4);
	
	\path [->] (v2) edge node[sloped,above,font=\scriptsize] {$e=0.49$} (v4);
	\path [dashed,line width=0.9pt,red,->] (v2) edge node[sloped,above,font=\scriptsize] {$e=0.47$} (v5);
	
	\path [->] (v3) edge node[sloped,above,font=\scriptsize] {$e=0.47$} (v5);
	\path [->] (v3) edge node[sloped,above,font=\scriptsize] {$e=0.45$} (v6);
	
	\path [->] (v4) edge node[sloped,above,font=\scriptsize] {$e=0.47$} (v5);
	\path [->] (v4) edge node[sloped,above,font=\scriptsize] {$e=0.45$} (v6);
	
	\path [->] (v5) edge node[sloped,above,font=\scriptsize] {$e=0.43$} (v7);
	\path [dashed,line width=0.9pt,red,->] (v5) edge node[sloped,above,font=\scriptsize] {$e=0.43$} (v9);
	
	\path [->] (v6) edge node[sloped,above,font=\scriptsize] {$e=0.43$} (v8);
	\path [->] (v9) edge node[sloped,above,font=\scriptsize] {$e=0.43$} (v6);
	
	\path [->] (v7) edge node[sloped,above,font=\scriptsize] {$e=0.45$} (v6);
	\path [->] (v7) edge node[sloped,above,font=\scriptsize] {$e=0.43$} (v9);
	
	\path [->] (v8) edge node[sloped,above,font=\scriptsize] {$e=0.43$} (v10);	
	\path [dashed,line width=0.9pt,red,->] (v9) edge node[sloped,above,font=\scriptsize] {$e=0.43$} (v10);
	
	\end{tikzpicture}
		\caption{} 
		\label{fig:AG}
	\end{subfigure} 
	\caption{(a) The cloud-model including 10 VMs in different hosts (servers). (b) Attack graph model corresponding to the cloud-model (capturing the connectivities of the VMs, note that the shortest attack path is shown as dashed lines).} 
	\label{fig:model}
\end{figure*}
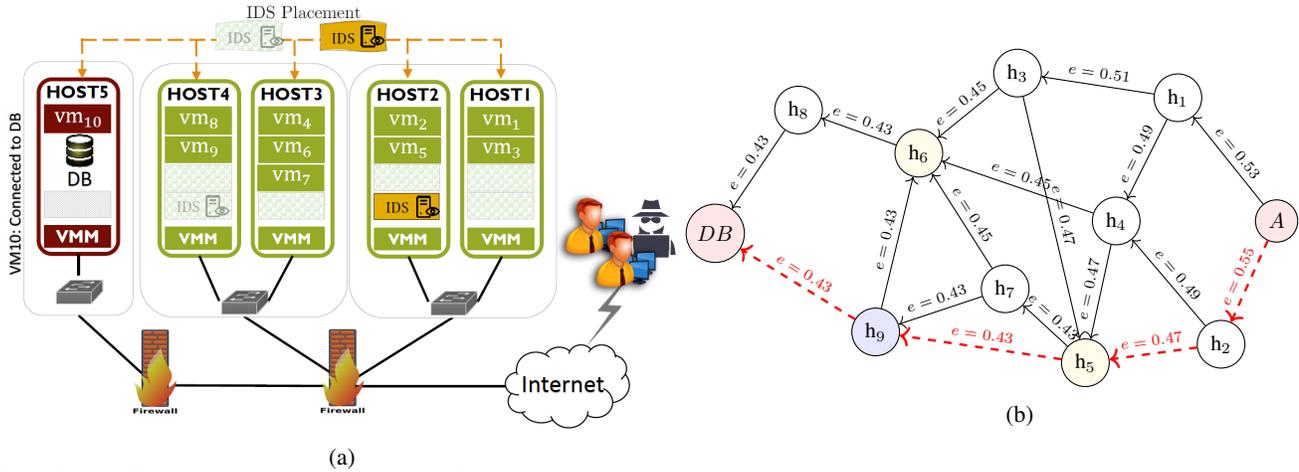 

We modeled a cloud system consisting of 10 Virtual Machines (VMs) distributed in 5 different physical servers or hosts in the cloud. We assume that only the VMs in host 1 (denoted as $h_1$) are connected to the internet and are the entry point of the system. The cloud model is demonstrated as in Figure~\ref{fig:model}. Each VM has a number of vulnerabilities associated with the Operating System (OS) it uses as in Table~\ref{vuls}. Thus, the attack model can be represented as a directed attack graph. {Let $AG = (V, E)$ be a graph, where 
$V$ is a set of all the nodes and $E$ is a set of all the edges. The aim of the attacker is to obtain the shortest attack path (SAP) which is a path between two nodes without considering the weight of the attack path. Note that the weight of the edges determines the exploitability $e$ of the connected VM based on Table~\ref{vuls}.}



\begin{table}[t]
	\centering
	\caption{Hosts vulnerabilities and exploitability information ($|V|$ is the number of vulnarabilities and Exploitability is the maximum exploitability of all vulnerabilities for a host)}
	\label{vuls}
	\footnotesize
	\begin{tabular}{|l|l|l|l|l|}
		\hline
		\multirow{2}{*}{VM} & \multirow{2}{*}{Current Host} & \multicolumn{2}{l|}{Vulnerabilities (V)} & \multirow{2}{*}{Impact} \\ \cline{3-4}
		&                                       & $|V|$             & Exploitability (e)               &                                          \\ \hline
		$vm_1$                    & $h_1$                             & 4                 & 0.53             & 10                                 \\ \hline
		$vm_2$                    & $h_2$                       & 4                 & 0.55             & 8                                 \\ \hline
		$vm_3$                    & $h_1$                               & 3                 & 0.51             & 9                                  \\ \hline
		$vm_4$                    & $h_3$                                & 3                 & 0.49             & 8                               \\ \hline
		$vm_5$                    & $h_2$                       & 2                 & 0.47             & 9                                \\ \hline
  	    $vm_6$                    & $h_3$                               & 1                 & 0.45             & 9                                \\ \hline
		$vm_7$                    & $h_3$                               & 1                 & 0.43             & 10                                \\ \hline
		$vm_8$                    & $h_4$                               & 1                 & 0.43             & 9                              \\ \hline
		$vm_9$                    & $h_4$                            & 1                 & 0.43             & 10                         \\ \hline
		$DB$                    & $h_5$                            & 1                 & 0.43             & 10                         \\ \hline
	\end{tabular}
\end{table}


\subsubsection{Attack Model}
{In this paper, we model an omniscient attacker which is able to traverse to reach the goal provided the perceived vulnerabilities exist.} {The attacker can launch AI-aided attacks using AI techniques to utilize multiple attack vectors by pursuing different exploits~\cite{zimba2019bayesian}. The attack usually comprises a sequence of transitions over time traversing the cloud network from one node to the others.} {As the attacker is an intelligent threat actor, he can use a variety of reconnaissance techniques to surveil the network for such vulnerabilities.}

\vspace{1mm}
\noindent{\textit{{General capabilities.}}}
We assume that the attacker is able to use various reconnaissance tools and techniques to gain enough information about the target system. The attacker has some information regarding the cloud hosts, VMs, and network gained through the various network and vulnerability scanning tools such as Nessus, Open Vulnerability Assessment Scanner (OpenVAS), etc. {The attacker can likewise infer adjacent networks by considering his subnet address and broadcast address. He further can have the topological overview of the associated subnets should he get hold of the corresponding routing tables.}

\vspace{1mm}
\noindent{\textit{{AI-based capabilities.}}}
We assume that the attacker is an AI-aided attacker which is able to leverage AI capabilities to gain valuable information about the system, targets, and attack paths. In this paper, we assume that the attacker can leverage AI to estimate the shortest attack path in the modeled cloud system. The shortest attack path for the cloud model is highlighted as a dashed line in Figure~\ref{fig:AG}. We assume that the attacker can leverage Deep learning Techniques (as in~\cite{salehi2020shortest}) to estimate the shortest path attack from the entry point of the system (i.e. internet) to the targeted host (DB) in the system. The attacker can exploit the vulnerabilities existing on each host with the probabilities defined based on the CVSS metrics. Note that for each attack step, the attacker needs to estimate the shortest attack path again as the system is dynamic and may be changed. However, for each attack step in an attack path, the attacker incurs some expenses such as costs and time. 


\vspace{1mm}
\noindent{\textit{{Attacker's goal and actions}}.}
The goal of the attacker is also well defined, which in this case is to exploit the database (DB) in the cloud through compromising the vulnerabilities existing on each VM in the attack path. 
The attacker has the ability to perform various actions. First, the attacker can take no action to hide. Once the attacker estimates the shortest attack path through AI techniques described in Section~\ref{sec:sp-DNN}, the attacker undertakes various actions to exploit the vulnerabilities of each VM in the attack path and finally exploit the target. Note that after exploiting a VM the attacker reaches a new state. We assume that the attack mounted by the attacker is {monotonic which means once an attacker has reached a certain state, they do not need to go back to any previous state, when targeting a specific goal.}

\vspace{1mm}

\begin{table*}[t]
\centering
\caption{Payoff matrix formalization based on $a^A, a^D$ for the game states $s_0$--$s_4$}
\label{TII}
\begin{tabular}{@{}p{2.1cm}p{2cm}ll@{}}
\toprule
\multicolumn{4}{c}{$s_0$: Initial State (no exploit)}                                    \\\midrule
A/D          & No-act          &  Def-$h_1$  &  Def-$h_2$  \\ \midrule
No-att & $0,0$ & $C_{def}, -C_{def}$ & $C_{def}, -C_{def}$  \\
$E(vm_1\in h_1)$  & $I_{vm_1},-I_{vm_1} $ & $-(I_{vm_1}-C_{def}), I_{vm_1}-C_{def}$ & $I_{vm_1}+C_{def}, -(I_{vm_1}+C_{def})$ \\
$E(vm_2 \in h_2)$ & $I_{vm_2},-I_{vm_2} $ & $I_{vm_2}+C_{def}, -(I_{vm_2}+C_{def})$ & $-(I_{vm_2}-C_{def}), I_{vm_2}-C_{def}$    \\\midrule
\multicolumn{4}{c}{$s_1$: Transition State ($vm_2 \in h_2$ exploited)}                                    \\\midrule
A/D          & No-act          &  Def-$h_3$  &  Def-$h_2$  \\ \midrule
No-att & $0,0$ & $C_{def}, -C_{def}$ & $C_{def}, -C_{def}$  \\
$E(vm_4 \in h_3)$  & $I_{vm_4},-I_{vm_4} $ & $-(I_{vm_4}-C_{def}), I_{vm_4}-C_{def}$ & $I_{vm_4}+C_{def}, -(I_{vm_4}+C_{def})$ \\
$E(vm_5 \in h_2)$ & $I_{vm_5},-I_{vm_5} $ & $I_{vm_5}+C_{def}, -(I_{vm_5}+C_{def})$ & $-(I_{vm_5}-C_{def}), I_{vm_5}-C_{def}$    \\\midrule
\multicolumn{4}{c}{$s_2$: Transition State ($vm_5 \in h_2$ exploited)}                                    \\\midrule
A/D          & No-act          &  Def-$h_3$  &  Def-$h_4$  \\ \midrule
No-att & $0,0$ & $C_{def}, -C_{def}$ & $C_{def}, -C_{def}$  \\
$E(vm_7 \in h_3)$  & $I_{vm_7},-I_{vm_7} $ & $-(I_{vm_7}-C_{def}), I_{vm_7}-C_{def}$ & $I_{vm_7}+C_{def}, -(I_{vm_7}+C_{def})$ \\
$E(vm_9\in h_4)$ & $I_{vm_9},-I_{vm_9} $ & $I_{vm_9}+C_{def}, -(I_{vm_9}+C_{def})$ & $-(I_{vm_9}-C_{def}), I_{vm_9}-C_{def}$    \\\midrule
\multicolumn{4}{c}{$s_3$: Transition State ($vm_9 \in h_4$ exploited)}          \\\midrule
A/D          & No-act          &  Def-$h_3$  &  Def-$h_5$  \\ \midrule
No-att & $0,0$ & $C_{def}, -C_{def}$ & $C_{def}, -C_{def}$  \\
$E(vm_6 \in h_3)$  & $I_{vm_6},-I_{vm_6} $ & $-(I_{vm_6}-C_{def}), I_{vm_6}-C_{def}$ & $I_{vm_6}+C_{def}, -(I_{vm_6}+C_{def})$ \\
$E(DB \in h_5)$ & $I_{DB},-I_{DB} $ & $I_{DB}+C_{def}, -(I_{DB}+C_{def})$ & $-(I_{DB}-C_{def}), I_{DB}-C_{def}$    \\\midrule
\multicolumn{4}{c}{$s_4$: Final State (DB exploited)}                             \\\midrule
\end{tabular}
\end{table*}

\subsection{Game Model Definition}
\subsubsection{Game Model Assumption}
We assume that the game for the defined cloud model defined in Section~\ref{sec:pre} can be modeled as a zero-sum Markov game in which the defender tries to place the IDS in the cloud's host to detect the attacker and defend against the attacker's action. These game attacks can be modeled using a Markovian model with finite states based on the attack. We also assume that both the players have full observability of the state in which they are. Additionally, we assume that the attacker can remain undetected in any VM in the cloud until it attempts to attack the other connected VM by exploiting the targeted VM vulnerabilities.

\subsubsection{Markovian Game Model}
{We model the attacker and a defender scenario as a two-player zero-sum Markov game leveraging the information in the cloud model and corresponding AG represented in Figure~\ref{fig:AG}. We also assume that the states and actions of the model are both discrete and finite. Moreover, the transition from each state and consequently  the corresponding reward for each state depends on players' actions for that specific state (this also can be modeled based on previous states and actions which are beyond the scope of this paper). We formally define a zero-sum Markov Game model based on obvious Markovian assumption and explain how each of these parameters are obtained in our cloud model.}
A Markov game for two players (in here, attacker and defender) can be defined by a tuple (S,$\mathcal{A}_A$, $\mathcal{A}_D$, T, R) where,

\begin{itemize}
    \item $S=\{s_0, s_1,s_2,\dots,s_r\}$ denoted the finite states of the game where in here $|S|=max(len(ap_i \in AP))$.
    \item $\mathcal{A}_A=\{a_0^A,a_1^A,a_2^A,...\}$ is the attacker's set of actions. The defender can have a set of actions  as $\mathcal{D}_A=\{a_0^D,a_1^D,a_2^D,...\}$. 
    \item $T=(s,a^A,a^D,s')$ is a States' Transition where the current state $s \in S$ will be changes to $s' \in S$ upon the actions come from both attacker and defender respectively. However, each transition has a probability which is denoted by $tp(T)$.
    \item $R^A(s;a^A; a^D)$ is the reward obtained by attacker if in state $s$, attacker and defender take the actions $a^A$ and $a^D$ respectively. However, the reward can be negative $-R^A(s; a^A; a^D)$ is the attacker choose a wrong action.
    \item $\lambda^p \in [0,1)$ is defined as the discount factor for the corresponding player $p$.
    \end{itemize}

\subsubsection{Reward function}
{The reward or payoff function depends on the actions taken by the attacker and the defender in each state of the game. We use the CVSS values defined in Table~\ref{vuls} for the reward or penalties associated with the successful or unsuccessful actions taken by either attacker or defender}.


To quantify reward values we use the important variables such as the impact of an attack and cost of defense ($C_{def}$), we used CVSS metrics that provide the Impact ($I$) for a specific VM ($I_{vm_i}$), Exploitability Scores ($e$), and other relevant metrics. $I_{vm_i}$ is a metric that computes the damage imposed to the VM by computing all impacts on the resources through an attack. For instance, $I_{vm_4}=8$ is the attack impact value on the VM $vm_4$ based on the related impact metrics of vulnerabilities in CVSS represented in Table~\ref{vuls}.
The rewards matrix for attackers is formulated as Equation~\ref{eq:reward_A}.

\begin{equation}
\small
\label{eq:reward_A}
R^\mathcal{A}_{a^A,a^D} \text{=}
  \begin{cases}
    0 &  \text{if}~a^A\subset \O\\
    C_{def}  &  \text{if}~a^A\subset\O,a^D\not\subset\O\\
    I_{vm_i}\text{+} C_{def}      &  \text{if}~a^A\text{=}E(vm_i),a^D\not\subset\O,vm_i\notin H(ids)\\
    I_{vm_i}  &  \text{if}~a^A\text{=}E(vm_i),a^D\subset\O\\
    \text{-}(I_{vm_i}\text{-} C_{def})      &  \text{if}~a^A\text{=}E(vm_i),a^D\not\subset\O,vm_i\in H(ids)\\
  \end{cases}
\end{equation}

Note that $H(ids)$ is a function that returns the host in which $ids$ has been located. For instance, if the defender locate the IDS in Host $h_4$, then $H(ids)$ returns $h_4$. As the game is a zero-sum game the reward for the defender is as equation~\ref{eq:reward_D}.
\begin{equation}
\label{eq:reward_D}
R^\mathcal{D}_{a^A,a^D}=-1~*~R^\mathcal{A}_{a^A,a^D}
\end{equation}

\begin{table}[t]
\centering
\caption{Payoff Matrix quantifying based on zero-sum game and CVSS values}
\label{TII-val}
\begin{tabular}{@{}p{2.1cm}p{2cm}ll@{}}
\toprule
\multicolumn{4}{c}{$s_0$: Initial State (no exploit)}                                    \\\midrule
A/D          & No-act          &  Def-$h_1$  &  Def-$h_2$  \\ \midrule
No-att & $0,0$ & $2, -2$ & $2, -2$  \\
$E(vm_1\in h_1)$  & $10,-10 $ & $-8, 8$ & $12,-12$ \\
$E(vm_2 \in h_2)$ & $8,-8 $ & $10,-10$ & $-6,6$    \\\midrule
\multicolumn{4}{c}{$s_1$: Transition State ($vm_2 \in h_2$ exploited)}                                    \\\midrule
A/D          & No-act          &  Def-$h_3$  &  Def-$h_2$  \\ \midrule
No-att & $0,0$ & $2, -2$ & $2, -2$  \\
$E(vm_4 \in h_3)$  & $8,-8 $ & $-6, 6$ & $10,-10$ \\
$E(vm_5 \in h_2)$ & $9,-9 $ & $11,-11$ & $-7,7$    \\\midrule
\multicolumn{4}{c}{$s_2$: Transition State ($vm_5 \in h_2$ exploited)}                                    \\\midrule
A/D          & No-act          &  Def-$h_3$  &  Def-$h_4$  \\ \midrule
No-att & $0,0$ & $2, -2$ & $2, -2$  \\
$E(vm_7 \in h_3)$  & $10,-10 $ & $-8,8$ & $12,-12$ \\
$E(vm_9\in h_4)$ & $10,-10 $ & $12,-12$ & $-8,8$    \\\midrule
\multicolumn{4}{c}{$s_3$: Transition State ($vm_9 \in h_4$ exploited)}          \\\midrule
A/D          & No-act          &  Def-$h_3$  &  Def-$h_5$  \\ \midrule
No-att & $0,0$ & $2, -2$ & $2, -2$  \\
$E(vm_6 \in h_3)$  & $9,-9$ & $-7,7$ & $11,-11$ \\
$E(DB \in h_5)$ & $10,-10 $ & $12,-12$ & $-8,8$   \\\bottomrule
\end{tabular}
\end{table}

As stated earlier, the formulation of the reward function is based on CVSS values and mainly the impact of the attack on a targeted VM. If the defender and the attacker do not take any action such that $a^A\subset\O,a^D\not\subset\O$ both get zero rewards. Moreover, if the attacker doesn't attack ($no$-$att$) while the defender place the IDS to any host in the cloud to secure any hosts, the defender incurs a cost for the defense ($-C_{def}$) and gets a negative reward. However, if the attacker attacks on a VM $vm_i$ while the defender place the IDS to detect attacks on the host in which the targeted VM $vm_i$ is located such that $vm_i\in H(ids)$, then the defender gets the reward for avoiding the attack impact on that VM ($I_{vm_i}$), but as the defender incurs some costs for the defense the total reward of successful defense is formulated as $I_{vm_i}-C_{def}$. For instance, suppose that the cost of defense is 2 units (for both successful or unsuccessful defense). Then, if attacker exploits VM $vm_1$ and defender put IDS on the host $h1$ ($vm_1\in h_1$), the defender gain a total reward of 7 which is as $R^D=I_{vm_1}-C_{def}=9-2$ while the attacker is penalized by -7 unit. In contrast, if the attacker attacks on a VM $vm_i$ while the defender place the IDS to detect attacks on the host in which the targeted VM $vm_i$ is not located such that $vm_i\notin H(ids)$, then the defender gets the penalty for wrong defense and incurs the impact of the attack on that VM plus the cost of wrong defense which is $-(I_{vm_i}+C_{ids})$ while the attacker reward would be as $I_{vm_i}+C_{ids}$ based on the zero-sum definition. For instance, if attacker exploits VM $vm_1$ and defender put IDS on the host $h2$ ($vm_1\notin h_2$), the defender gets a negative reward of -11 which is as the sum of the impact of attack on that VM and the cost of defense as $R^D=-1*(I_{vm_1}+C_{def})=-1*(9+2)$. Then, the attacker gets rewards of 11 which is $R^A=-R^D$. Lastly, if the attacker attacks on a VM and the defender takes no action then the attacker gains the reward for the successful attack which is equivalent to the impact of the attack on exploited VM $vm_i$ as $R^A=I_{vm_i}$ while the defender gets a negative reward as $R^D=-I_{vm_i}$.

A normal-form zero-sum reward matrix for the four states of the game in the Markov game is shown in Table~\ref{TII-val} which is quantified based on the CVSS values and the reward function formulation explained before.



\begin{figure*}[t]
	\centering
	\includegraphics[width=0.90\linewidth]{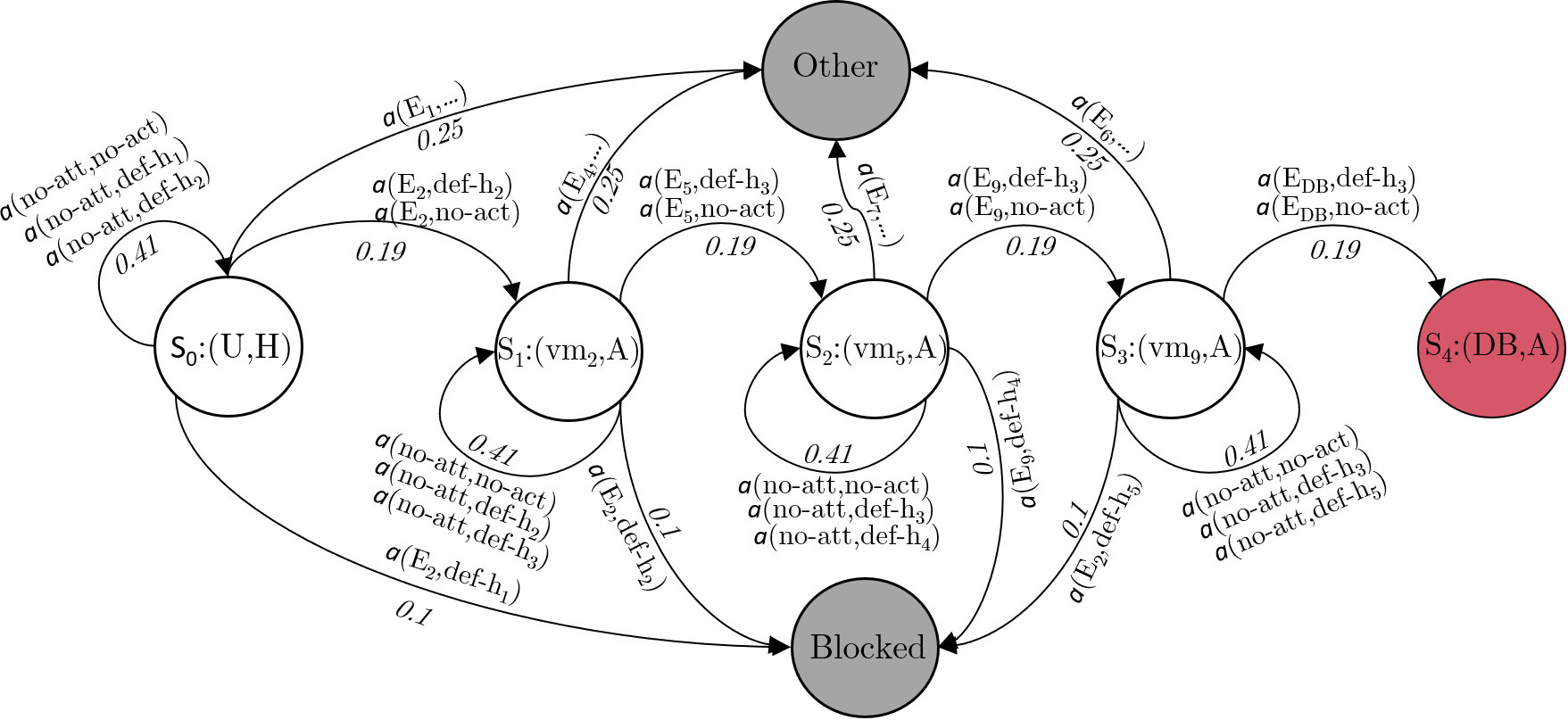}
	\caption{Markov model of the game with finite states and deterministic probabilities based on the shortest attack path.}
	\label{fig:MK}
\end{figure*}

\subsubsection{States, actions and transitions}
The Markov model of the proposed game is illustrated in Figure~\ref{fig:MK} which captures the transitions and associated probabilities in which the attacker tries to find the shortest attack path to exploit DB (based on Figure~\ref{fig:net}).

\vspace{1mm}
\noindent{\textit{{States}}.}
{It represents the state attacker/defender currently have in the cloud over different preformed actions. We extract the information from the shortest path in the cloud attack graph to define the states. For instance, for the attacker, initial state $s_0 = (Host; User)$, if the successful execution of the exploit of VM $vm_2$ is performed by the attacker $E(vm_2)$, the attacker can transition to another state $s_1 =  (H_1; Attacker)$.}

\vspace{1mm}
\noindent{\textit{{Actions and state transitions}.}}
{Based on the system model represented in Figure~\ref{fig:AG}, the attacker has at most three possible actions in each state. The attacker can choose no attack ($no$-$att$ or $\O$) or attack to another adjacent VM by exploiting the vulnerabilities of targeted VM. Thus, for each state the maximum actions can be defined as $Max(Deg(vm_i \in H) + 1 = 3$.
For instance, in $s_0$, the action space for the attacker can be as $a^A_{0,s_0} =\O$, $a^A_{1,s_0} =E(vm_1)$, $a^A_{2,s_0}=E(vm_2)$. Similarly, the defender has its own possible actions to defend ($Def$) hosts. For instance, the defender can perform no defence ($No$-$act$ or $\O$). All possible actions for the defender in state $s_0$ is as $a^D_{0,s_0} = \O$, $a^D_{1,s_0} = D(h_1)$, $a^D_{2,s_0} = D(h_2)$.}

\subsection{Probabilistic Model and Game Solving}
\subsubsection{Uniform Random Strategy (URS)} 
We assume that the defender uses Uniform Random Strategy (URS) where the defender selects the actions $a^D_s \in \mathcal{D}_A$ based on a uniform probability distribution over its possible actions in the corresponding state. The decision-making process can be viewed as randomization that chooses the next valid state based on a specific probability distribution over states $s_q \in S$, then choosing the next host for IDS placement from a uniform distribution to become specific instances of randomization. We choose URS as the baseline of the defender's strategy. For instance, based on the initial state $s_0$
shown in Table~\ref{TII-val}, The defender can select the mixed strategy
of placing IDS on host $h_1$, host $h_2$ in the cloud, or taking no action. Thus, the selection of actions is uniformly distributed for the defender with the equal probability of $3.33$. However, many studies claim that defender's strategy selection can be performed as pure or URS for dynamic defense~\cite{zhuang2014towards}. 

\subsubsection{Maxmin strategy}
In this game, the attacker $P_A$ aims to maximize his expected discounted reward and the defender $P_D$ tries to choose the actions that minimize the expected reward for the attacker. Thus, the maxmin strategy can be considered for calculating the expected reward of $P_A$ in the Markov game.
{Given $Q(s,a^A, a^D)$, an agent is able to maximize the reward using the greedy strategy by always selecting the action having the highest $Q$-value. This strategy is considered as a greedy strategy because it treats $Q(s,a^A, a^D)$ as a surrogate for immediate reward and then acts to maximize its immediate gain. However, it is optimal because the $Q$-function is an accurate summary of future rewards. }

We now define the quality of an action or the $Q$ value used to represent the expected reward the attacker (A) will get for choosing the action $a^A \in \mathcal{A}_A$ while the defender chooses $a^D \in \mathcal{D}_A$. 

$$
Q(s,a^A, a^D)=R(s,a^A, a^D)+\lambda \sum_{s'} T(s,a^A,a^D,s')
$$
While the value of a state $s \in S$ in a Markov game is as

$$
V(s)= max_{\pi(s)}~min_{a^D}\sum_{a^A} Q(s,a^A,a^D).\pi_{a^A}
$$
However, the defender tries to minimize the attacker's reward by placing IDS in the various hosts in the cloud that are a part of the attack paths in the attack graph, while the attacker aims to use a mixed policy $\pi(s)$ over it possible actions in $a^D$ to maximize its total reward. Thus, the Markov Game is useful framework for the defender to model the attacker's policy so that they can take necessary actions and countermeasures by making decision in each state to minimize the expected attacker's utility.

\subsubsection{Transitions Probabilities Assignment} 
The actions for attackers and defender are considered separately for each states. For instance, the attacker action space in the initial state $S_0$ is as $\mathcal{A}_{A,s_0}=\{a^A_{0,s_0},a^A_{1,s_0},a^A_{2,s_0}\}$ where $a^A_{0,s_0}=\O$ which indicates the attacker takes no action/attack ($No$-$att$) to avoid detection, $a^A_{1,s_0}=E_1$ which implies that the attacker exploits VM $vm_1$ (note that $E(vm_i)$ is shortly denoted as $E_i$), and $a^A_{2,s_0}=E_2$ which means exploiting of VM $vm_2$ or $E(vm_2)$. The probability of attack access through considering all possible actions for each state $s_j$, denoted as $p(AS_{s_j})$, can be defined as the Equation~\eqref{eq:pas}. This means the attacker can launch a successful attack by taking only one successful action in that state (note that the action $No$-$att$ is not considered as a successful attack action).

\begin{equation}
\label{eq:pas}
p(AS)_{s_q}= 1-\prod_{a^A_{j,s_q} \in A_{A,s_q}-\{\O\}}{\Big(1-e(a^A_{j,s_q})\Big)}
\end{equation}

Note that $e(a^A_{j,s_q})$ is the probability of attack success by taking the specific action $a^A_{j}$ in a state $s_q$ which is the exploitability of the targeted VM based on Table~\ref{vuls}. For instance, exploiting of $vm_1$ is an action of the attacker $a^A_{1,s_0}=E_1$, then $e(a^A_{1,s_0})$ is $e(E_1)=e(vm_1)=0.53$. 

Now we define the probability that the attacker chooses a specific action $a^A_z$ in a current state $S_q$ as Equation~\eqref{eq:ps}.

\begin{equation}
\label{eq:ps}
p(a^A_{z,s_q})=\frac{e(a^A_{z,s_q})}{\sum_{a^A_{j,s_q} \in A_{A,s_q}}{\Big(e(a^A_{j,s_q})\Big)}}
\end{equation}

For instance, the probability that attacker takes action $a^A_{1,s_0}$ in state $s_0$ which means that the attacker prefers to exploit $vm_1$ (denoted as $p(a^A_{1,s_0})$ or $p(E_1)$) is calculated as:

$$
p(a^A_{1,s_0})=\frac{e(a^A_{1,s_0})}{e(a^A_{1,s_0})+e(a^A_{2,s_0})}=\frac{e(E_1)}{e(E_1)+e(E_2)}
$$

Based on the above equation, the result of $p(E_1)$ is as $\frac{0.53}{1.08}\approx 0.49$. Similarly, the probability of the attacker choose the second action $p(E_2)$ is computed as $p(a^A_{2,s_0})=p(E_2)\approx 0.51$.

We then define the transition probability for attackers only for a specific attack action ($a^A_z$) as Equation~\eqref{eq:TA}. 

\begin{equation}
\label{eq:TA}
\tau(a^A_{z,s_q}) =
  \begin{cases}
    p(a^A_{z,s_q}).e(z,s_q) \hspace{25mm} \small \text{if}~a^A_{z,sq}\not\subset\O\\
    1-\sum_{a^A_{j,s_q} \in A_{A,s_q}}{\Big( p(a^A_{j,s_q}).e(j,s_q) \Big)} ~~  \small \text{otherwise}\\
  \end{cases}
\end{equation}

For instance, the $\tau(a^A_{2,s_0})= \tau (E_2)= 0.51*0.55=0.28$ which is the product of the probability that the attacker choose action $a^A_{2,s_0}$ and the attack success probability of the related attack action $e(a^A_{2,s_0})$. We then assume that the probability of defender's actions for each state of the same are uniformly distributed. Thus, the transition probability for defender only for n specific defend action ($a^D_z$) for the state $s_q$ is defined as:

\begin{equation}
\label{eq:TA}
\tau(a^D_{z,s_q}) =\frac{1}{|\mathcal{D}_{A,s_q}|},
\end{equation}
where $|\mathcal{D}_{A,s_q}|$ is the numbers of actions for defender. For instance, if the defender has three actions such as no action ($no$-$act$), defend host $h_1$, and defend host $h_2$, then $\tau(a^D_{2,s_0})=\tau(D_2)\approx 0.33$. Note that $D_2 \in \mathcal{D}_A$ indicates the defend of host $h_2$ (placement of IDS in host $h_2$).

Now we define the transaction probability based on both attacker's and defender's actions as Equation~\eqref{eq:T}.


\begin{equation}
\label{eq:T}
tp(s,a^A_z,a^D_z,s')=\tau(a^A_{z,s}).\tau(a^D_{z,s})
\end{equation}

For example, the transition probability for $T_{0,1} = (s_0,a^A,a^D,s_1)$ can be computed as:
$$
tp(T_{0,1})=tp(s_0,a^A,a^D,s_1)=\tau(E_2).\tau(D_2) \cup \tau(E_2).\tau(\O)  \\ 
$$
which yields $p(T_{0,1}) \approx 0.19$. Similarly, the transition from state $s_0$ to $s_0$ can be defined as $T_{0,0} = (s_0,a^A,a^D,s_0)$ and its probability is computed as:

\begin{equation}\label{optimization_model}
\begin{split}
tp(T_{0,0})=tp(s_0,a^A,a^D,s_0)=\tau(\O).\tau(\O) \cup \tau(\O).\tau(D_1)  \\
\cup \tau(\O).\tau(D_2)\approx 0.41.
\end{split}
\end{equation}
Likewise, all the Markovian model transitions for the states of the game and the transition probabilities are computed and illustrated in Figure~\ref{fig:MK}. Note that the Markovian game is modeled based on the attacker using the shortest attack path to find the target for each state of the game and the probabilities are defined and formulated based on the URS for this game. However, transitions probabilities assignment though Q-learning is not considered in this paper and will be considered in our future work.



\section{Discussions and Limitations}\label{sec:discussion}
This paper attempts to initiate a discussion regarding game theory evolution which can be able to evaluate AI-aided threats and making appropriate decisions on possible defensive strategies. In this paper, we only considered an example of AI-aided attacks which is able to leverage the deep neural networks to find the shortest attack path in a networked system. However, a more capable game model needs to be proposed to be able to defend against AI-embedded attacks such as Deep locker. We summarize some limitation and further challenges as follows:

\vspace{1mm}
\noindent{\textit{Limitation and Challenges}.} Following challenges need to be investigated more in the current game theory models against the next generation of cyber threats. 

\begin{figure}[t]
	\centering
	\includegraphics[width=0.85\linewidth]{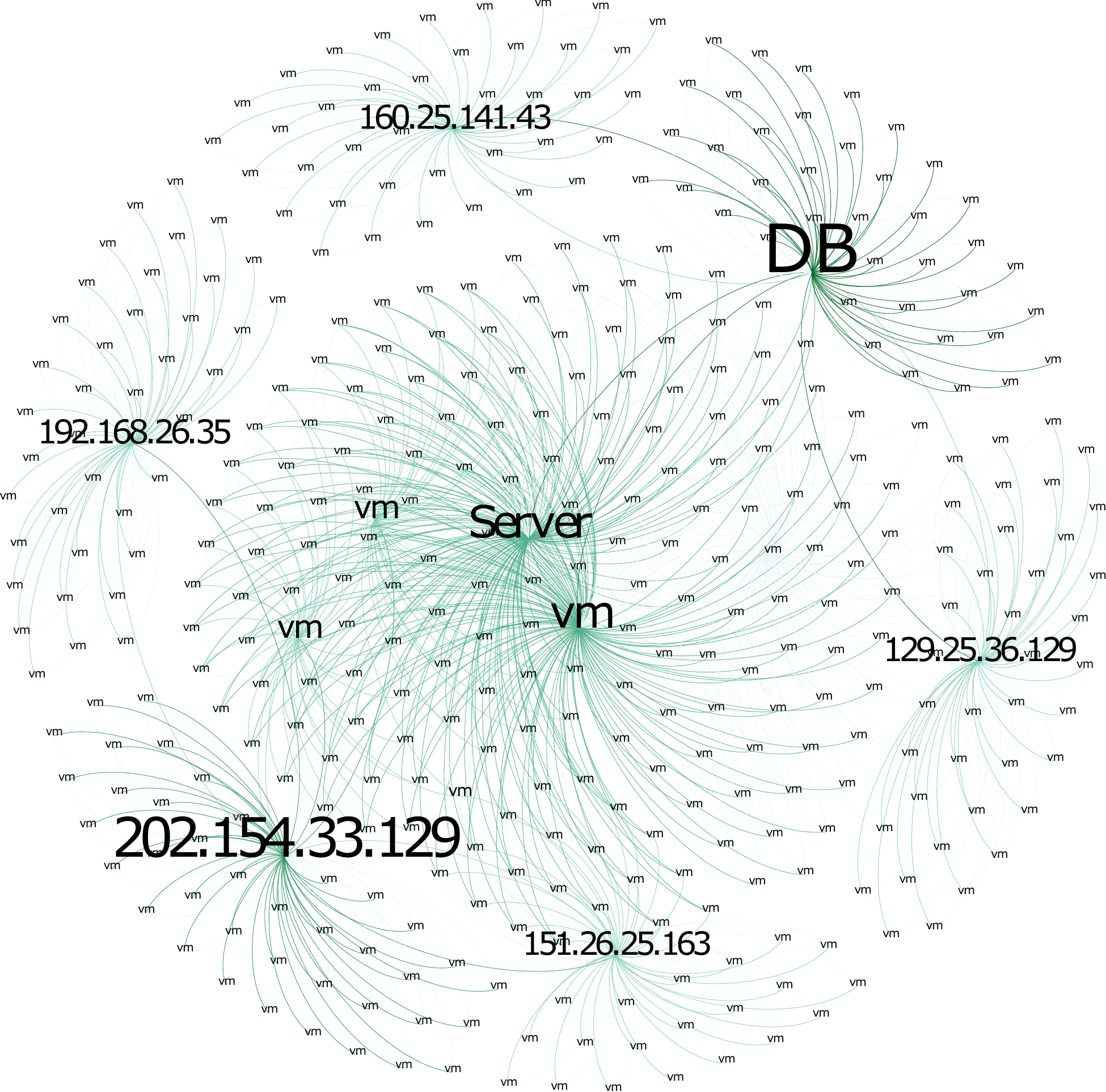}
	\caption{AG generated for a large cloud model. The AI-aided attacker can leverage DNN to estimate the shortest attack path effectively in large cloud model.}
	\label{fig:AG-Large}
\end{figure}

\begin{itemize}
	\item In this paper, we only modeled an attacker that is able to estimate the shortest attack path in the modeled cloud. However, other efficacy factors need also be considered in the model such as maximum exploitability ($ME$). It can ensure that the attacker traverses the attack paths which yields the maximum probability of success. The omniscient can leverage AI techniques as defined in Section~\ref{sec:sp-DNN} and \cite{zimba2019bayesian} to estimate and probe not only the shortest path but also the best attack path in the network. Thus, the attacker can find the attack path with the maximum exploitability values based on CVSS values. In fact, the attacker goes through the path(s) with the highest attack success probability values. Based on the attacker's point of view, $ME$ can be defined as the following equation.
$$
ME=\min_{ap \in {AP}}\Big(\sum_{vm_i \in ap} e({vm_i})\Big)
$$
However, in this paper, the main focus is on the attacker using DNN to find the shortest attack path and we aim to further consider other attack strategies such as $ME$ estimation in our future work.
	\item We only considered our proposed game theory model for a small cloud model. However, this can be extended to larger cloud models in which the attacker can find the shortest attack path through DNN more efficiently. The large cloud system can be modeled using an AG,~see Figure~\ref{fig:AG-Large}. In this case, the maximum number of states ($S$) in the game model can is determined as the maximum number of VMs in the shortest attack path: 
    $$
    SAP = min\big(len(ap_i \in AP)\big),
    $$
	then, $|S|=SAP$
	As the model is based on the shortest attack path, the model is not faced with a state explosion problem.
    \item An important limitation of the game-theoretic approach in cyber security is the difficulty to perfectly quantifying parameters of cyberspace. However, this might affect the decision-making process by defenders. However, it is important to leverage Machine learning-based approaches to identify parameters and values for both attacker and defender. 
	\item The application of Nash equilibrium requires both attackers and defenders to choose their own optimal strategies at the same time which the process is difficult to be achieved in the reality especially for the new generation of threats.
	\item However, in most game theory models, the investigations are based on the hypothesis for both sides of attacker and defender and are completely rational. Both parties know how to realize the maximization of their reward values. However, the attack-defense information of the actual network is  and intimate and asymmetric as the strategies' rewards could be private information for the game players.
\end{itemize}

\section{Conclusion}\label{sec:conclusion} 
This paper first discusses the different types of AI-powered attacks categorized as AI-aided and AI-embedded attacks. Then the application of game theory to model various cyber threat scenarios is reviewed. Then, a cloud system is proposed in which an AI-aided attacker can find the shortest attack path in the network effectively using a deep neural network. We proposed a zero-sum Markovian game model which is able to model AI-aided attacks based on finite states which can help the defender to make appropriate decision to mitigate the attack impact. This paper demonstrates the potential of Markovian game theory models in strengthening the decision-making based on the capabilities of AI-based threats and finding optimized strategies in decision making comparing with other time-consuming optimization techniques. However, there are several critical challenges that need to be further studied and addressed before modeling the AI-based game theory models. We hope that our discussion and our initial design of an experimental model can trigger more profound research to make the game theory more viable for novel cyber threat evaluation.

\bibliographystyle{IEEEtran}
\bibliography{IEEEabrv,GT-attack}

\end{document}